\documentclass[aip,jcp,reprint,amssymb,amsmath,superscriptaddress,groupedaddress,frontmatterverbose,]{revtex4-2}

\usepackage[title]{appendix}
\usepackage{graphicx, tikz, orcidlink}
\usepackage{epstopdf}
\usepackage{ascmac}
\usepackage{ulem}
\usepackage{cancel}
\usepackage{here}
\usepackage{physics}
\usepackage{bm}
\usepackage{comment}
\usepackage{diagbox}
\usepackage{color}

\begin{document}

\title{MO-HEOM: Extending Hierarchical Equations of Motion to Molecular Orbital Space}
\date{Last updated: \today}

\author{Yankai Zhang\orcidlink{0009-0002-0439-2817}}
\author{Yoshitaka Tanimura\orcidlink{0000-0002-7913-054X}}
\email[Author to whom correspondence should be addressed: ]{tanimura.yoshitaka.5w@kyoto-u.jp}
\affiliation{Department of Chemistry, Graduate School of Science,
Kyoto University, Kyoto 606-8502, Japan}
\author{So Hirata\orcidlink{0000-0002-9578-1502}}\affiliation{Department of Chemistry, University of Illinois at Urbana-Champaign, Urbana, IL, USA}

\begin{abstract}
Studies of quantum thermal effects on molecular excitation dynamics have traditionally relied on oversimplified models--energy eigenstates or low-dimensional potentials--that inadequately represent the complexity of real chemical systems. In practice, molecules are spatially extended and embedded in anisotropic environments, where molecular orbitals (MOs) decisively govern quantum behavior. To advance beyond these limitations, we propose a three-dimensional rotationally invariant system–bath (3D-RISB) model within the MO framework, with explicit inclusion of intramolecular vibrational motion. From this MO foundation, we derive numerically ``exact'' hierarchical equations of motion (MO-HEOM). As a demonstration, we analyze hydrogen molecules and hydrogen molecular ions with vibrational degrees of freedom, revealing their linear absorption spectra.
\end{abstract}

\maketitle

\section{Introduction}\label{sec:intro}
The essence of theoretical chemistry resides in the interplay between quantum chemistry and statistical thermodynamics. Together, these disciplines form a unified framework that underpins our understanding of chemical phenomena.
Quantum chemistry is grounded in microscopic equations that govern the behavior of atoms and electrons.\cite{LCAO1931,HMO1931,SCF1951,SCFMETHOD1965,DFT1965,Nakatsuji1978,Nakatsuji1983} These equations allow for the prediction and analysis of molecular structures that emerge from chemical bonding. Modern quantum chemistry enables the theoretical design of molecules with targeted structures and properties, even in the absence of experimental data. Thermodynamics, on the other hand, provides a macroscopic description of energy and matter, while statistical mechanics offers a many-body formulation of thermal equilibrium.\cite{Oono2017} 
This formulation opens up the possibility of defining fundamental concepts, including temperature and entropy, rigorously from first principles.\cite{KT24JCP1,KT24JCP2,EspositoHarbolaMukamel2009}

To elucidate the behavior of chemical systems, both microscopic and macroscopic perspectives are indispensable. For instance, many chemical reactions occur exclusively in solution or within biological environments, indicating that reaction dynamics cannot be fully captured by potential energy surfaces derived from few-body quantum models. Without accounting for thermal activation, energy redistribution, and the diffusion of excess energy, effects that arise from interactions with a large or effectively infinite number of degrees of freedom, accurate prediction becomes unattainable.

Yet, reconciling the microscopic formalism of quantum chemistry with the macroscopic nature of statistical thermodynamics remains a formidable challenge. Conventional reaction rate theories incorporate temperature-dependent constants but do not address the origin of thermal energy, rendering them qualitative and limited in predictive power. Dynamical approaches such as surface hopping rely on phenomenological approximations and neglect a quantum treatment of the thermal bath. As a result, their validity is confined to regimes that remain close to classical behavior.\cite{IT19JCTC}

Recent advances in quantum thermodynamics\cite{KT24JCP1,KT24JCP2,GemmerMahlerMichel2004,EspositoHarbolaMukamel2009,RivasHuelgaAlonso2014,GooldHuberRieraDelRioSkrzypczyk2016,MenczelFunoCirioLambertNori2024} and quantum information science,\cite{NielsenChuang2000,Shor1994,BennettBrassard1984,NationJohanssonBlencoweNori2012} which focus on simple systems such as atoms, spins, and nanomaterials, provide valuable insights into thermal processes. A prototypical model involves a two-level system\cite{SpinBosonLeggett} or a one-dimensional potential system\cite{CALDEIRA1983587,GRABERT1988115} 
 coupled to a bath described by a continuum of harmonic oscillators, as a thermal bath. These oscillators, distributed canonically, behave as a bath with infinite specific heat and follow microcanonical statistics. Upon time evolution, the system and bath become quantum mechanically entangled (``bathentanglement''),\cite{T20JCP} reaching a thermal equilibrium state.

A key discovery in this context is the inherently non-Markovian nature of quantum thermal noise, characterized by Matsubara frequencies.\cite{T06JPSJ,T20JCP} Even when the interaction between the system and bath is weak, repeated interactions within the noise correlation time necessitate a non-perturbative treatment.\cite{KT24JCP3,KT24JCP4}

In numerical simulations, the infinite degrees of freedom of the bath cannot be treated explicitly. Instead, they are integrated out using canonical weights, yielding a reduced density matrix whose time evolution is governed by the theory of open quantum dynamics. While the Markov approximation leads to semiclassical results valid only at high temperatures, non-Markovian and non-perturbative approaches such as the hierarchical equations of motion (HEOM)\cite{TK89JPSJ1,IT05JPSJ,T90PRA,TW91PRA,TW92JCP,T06JPSJ,T20JCP,T20JCP,KT24JCP3,KT24JCP4}
 and the quasi-adiabatic path integral (QUAPI) method\cite{Makri95, Makri96, Makri96B}  provide rigorous and quantitative descriptions of thermal environments. 

With the development of advanced algorithms and high-performance computing, it is now possible to simulate large quantum systems with remarkable precision.\cite{Burghardt2011,Plenio2018,Lambert2019,Petruccione2020,pseudomodesNori2024,Shi2009HEOM,SHI10.1063/1.5026753,IKEDASCHOLES2020,IKEDANAKAYAMA2022,Huang2023,Ke2023TreeTensorHEOM,Borrelli2021,Takahashi2024,Ignacio2024,Ignacio2025,LinjunWan2025}  However, when applying dissipative quantum mechanics to molecular orbital theory, a critical condition must be met: the system must retain its symmetry to preserve quantum entanglement with the bath.

This requirement emerged from studies of quantum rotors. Classically, rotational relaxation is described by Langevin dynamics. In the quantum regime, it is modeled by a rotational Hamiltonian coupled to a harmonic oscillator bath. Surprisingly, quantum rotational transitions vanish from the absorption spectrum regardless of temperature.\cite{ST01JPSJ} This phenomenon arises because the standard bath lacks the symmetry of the rotational system, leading to the destruction of entanglement and a return to classical behavior.

This issue was resolved by introducing a spatially structured bath that interacts with the system in a symmetry-preserving manner. The rotationally invariant system-bath (RISB) model, originally developed by Gefen, Ben-Jacob, and Caldeira\cite{1987RISBPhysRevB.36.2770}  for tunnel junctions, demonstrated temperature-dependent linewidths in rotational spectra\cite{IT18JCP}  and revealed off-diagonal peaks in two-dimensional (2D) terahertz spectroscopy that originate from the selection rules.\cite{IT19JCP,LipenRISB_HEOM}

Building upon this foundation, a recent investigation utilized a hierarchical equations of motion (HEOM) constructed from the 3D RISB framework to explore the Coulomb potential in terms of atomic orbitals. This theoretical framework, AO-HEOM, allows for a rigorous treatment of quantum dynamics in atomic systems, explicitly accounting for thermal fluctuations and dissipative processes.\cite{ZT25JCP1,ZT26JCP1} For example, in the hydrogen atom and the helium ion (He$^+$), the spectral features of the Lyman and Balmer series have been shown to vary with temperature and with the strength of thermal coupling. It has further been demonstrated for the absorption spectrum of hydrogen atom that transitions from high-energy states are suppressed by strong rotational relaxation, even when such transitions would otherwise be allowed by the selection rules.\cite{ZT25JCP1}

The present work extends this approach from atomic to molecular orbitals derived from quantum chemical calculations. To accommodate molecular symmetry, the bath is generalized to 3D, allowing for anisotropic environmental effects. As a demonstration, we investigate hydrogen molecule and hydrogen molecule ion coupled to a bath modeled by the 3D RISB framework. To validate the theory, we compute the linear absorption spectra.

This paper is structured as follows. Section \ref{sec:Model} introduces the model system for molecular electronic states. The MO-HEOM are then presented in Sec. \ref{sec:MO-HEOM}.
Section \ref{sec:NumericalDemo} provides numerical demonstrations. Section \ref{sec:Conclution} concludes with a summary and outlook.

\section{Molecular electronic systems in dissipative environments}
\label{sec:Model}

\subsection{Electronic Hamiltonian for a Diatomic Molecule}

Our ultimate goal is to describe the dynamic properties of electronic states in diverse molecular systems within the framework of open quantum dynamics theory. However, to clarify the key points of the discussion, it is preferable to begin with a simple model.
Consequently, we begin by examining a diatomic molecule with nuclear charges denoted by $Z_A$ and $Z_B$, 
which serves as a simple prototype for our theoretical framework.
The coordinates of the nuclei are denoted by ${\bf R}_A$ and ${\bf R}_B$, respectively, and the interatomic distance is defined as
$r=|{\bf R}_A-{\bf R}_B|$.
The molecular Hamiltonian for a diatomic molecule is expressed as
\begin{align}
\hat{H}_\mathrm{S} = \hat{H}_e (r) + \hat{H}_v,
\label{eq:H2}
\end{align}
where 
\begin{align}
\hat{H}_e (r) &=- \sum_{i}^N \left[ \frac{\hbar^2}{2m_e} \nabla_i^2 + \frac{e^2}{4\pi\varepsilon_0} \left( \frac{Z_A}{r_{iA}} + \frac{Z_B}{r_{iB}} \right)    \right] \nonumber \\
& + \sum_{i<j}^{N} \frac{e^2}{4\pi\varepsilon_0} \cdot \frac{1}{r_{ij}} 
\label{eq:MO}
\end{align}
and $m_e$ is the mass of the electron, $\nabla_i^2$ is the Laplacian operator acting on electron $i$  representing its kinetic energy, $e$ is the elementary charge, $\varepsilon_0$ is the vacuum permittivity. We definehe distances between electron $i$ and nucleus $A$ and $B$ as
$r_{i A} = |\mathbf{r}_i - \mathbf{R}_{A} |$ and $r_{i B} = |\mathbf{r}_i - \mathbf{R}_{B} |$, respectively. 
The inter-electronic distance is given by $r_{ij} = |\mathbf{r}_i - \mathbf{r}_j|$.
The Hamiltonian of the nuclei is expressed as
\begin{align}
\hat{H}_v = - \frac{\hat{p}^2}{2\mu}  + U(\hat  r),
\label{eq:nuclei}
\end{align}
where $\hat{p}$ is the conjugate momentum operator with respect to the nuclear coordinates $\hat r$, $\mu$ denotes the reduced mass, and $U(\hat r)$ represents the potential energy associated with nuclear motion.

The dipole operator of a molecular system consists of electronic and nuclear contributions:
\begin{align}
\hat{\bm{\mu}} = -e \sum_{i=1}^N \bm{r}_i +  e Z_A  \bm{R}_A +  e Z_B \bm{R}_B
\end{align}

For the diatomic  molecule, we set the $z$-direction as the principal axis of the hydrogen molecule. 
We denote the equilibrium position of the electronic ground state by $r^{eq}$.  
For a displacement ${q} = {r} - r^{\mathrm{eq}}$, the diabatic potential
surface of the $n$th electronic state under the Born--Oppenheimer approximation
is given by
\begin{equation}
    U_{nn}(q) = \langle n, q \rvert U(\hat{q}) \lvert n, q \rangle ,
\end{equation}
where $\lvert n, q \rangle$ denotes the $n$th electronic eigenstate as a
function of the nuclear coordinate, obtained from MO calculations.
The off-diagonal element $U_{nm}(q)$ represents the nonadiabatic coupling
between the $n$th and $m$th states, which is also evaluated from MO
calculations and arises from the breakdown of the Born--Oppenheimer
approximation.

The Hamiltonian of the system is defined as\cite{mukamel1999principles,TM94JCP,TM97JCP,MT98CPL,IT17JCP,IT18CP,IT19JCTC,IDT19JCP}
\begin{align}
  \hat{\bm{H}}_{\mathrm{S}} &= \frac{\hat{p}^{2}}{2\mu} \otimes \bm{I} + \sum_{n,m} |n\rangle U_{nm}(\hat{q}) \langle m|, 
  \label{eq:system-Hamiltonian}
\end{align}
where $\bm{I} \equiv \sum_{n} |n\rangle \langle n|$.  
For the multi-state system described by Eq.\ \eqref{eq:system-Hamiltonian}, the density matrix can be expanded in the electronic basis as
\begin{align}
  \hat{\bm{\rho}}(q, q'; t) &= \sum_{n,m} |n\rangle \rho_{nm}(q, q'; t) \langle m|, 
  \label{eq:system-densitymatrix}
\end{align}
where $\rho_{nm}(q, q'; t)$ is the density matrix element for the electronic states $n$ and $m$ in the coordinate representation.

\subsection{Linear absorption spectrum}

To calculate the spectrum, the transition dipole operator, expressed as
\begin{eqnarray}
\hat{\mu}_{\alpha} = \sum_n \mu_{n0}^{\alpha}(q) \lvert n \rangle \langle 0 \rvert + h.c.,
\label{eq:mu}
\end{eqnarray}
must also be evaluated. In the case of a diatomic molecule, while the
$z$-component of the transition dipole between the ground state and the
$n$th excited state depends on $q$, the $x$ and $y$ components are
independent of the nuclear coordinate.

We demonstrate our formalism by depicting linear absorption spectrum, defined as:\cite{IT19JCP}
\begin{eqnarray}
 &&I_{\alpha' \alpha }(\omega)= \mathrm{Im}\left(\frac{i}{\hbar }\right)\int _{0}^{\infty }\mathrm{d}t e^{i\omega _{1}t_{1}} 
\mathrm{Tr}\left\{\hat{\mu}_{\alpha'}\hat{\mathcal{G}}(t)\hat{\mu }_{\alpha}^{\times }\hat{\rho}_{\mathrm{eq}}\right\}, 
  \label{eq:R1}
\end{eqnarray}
where we introduce the hyperoperators $\hat{A}^{\times} \hat{B} \equiv \hat{A} \hat{B} - \hat{B} \hat{A}$, defined for arbitrary operators $\hat{A}$ and $\hat{B}$, and $\hat{\mathcal{G}}(t)$ is Green's function in the absence of a laser interaction.

\subsection{Effects of environments}

Previous studies for AO-HEOM\cite{ZT25JCP1,ZT26JCP1} have primarily focused on thermal baths of electromagnetic origin. However, in molecular systems, the relaxation of molecular vibrations through interactions with the condensed phase--such as the solvent--becomes important in addition to electromagnetic interactions. This paper incorporates both types of relaxation into the formalism.

The total Hamiltonian including the baths for electromagnetic and mlecular vibrational relaxation is expressed as
\begin{align}
  \hat{\bm{H}}_{\mathrm{tot}} &= \hat{\bm{H}}_{\mathrm{S}} + \hat{\bm{H}}_{\mathrm{I+B}}^v  + \hat{\bm{H}}_{\mathrm{I+B}}^e.
  \label{eq:total-Hamiltonian}
\end{align}
The detailed forms of each thermal bath are described in the following subsections.

\subsubsection{Molecular vibrational relaxation}
\label{sec:MO}
When analyzing relaxation in various vibrational modes of molecules with 3D structures, each mode must be decomposed into its coordinate components, and relaxation should be treated within the framework of the 3D-RISB model.
However, since we are considering a diatomic molecule here, we adopt a model in which a single thermal bath interacts along the principal vibrational axis.

The bath Hamiltonian in this case can be expressed in the following form.
\begin{align}
  \hat{\bm{H}}_{\mathrm{I+B}}^v &= \sum_{j} \left\{ \frac{\hat{p}_{j}^{2}}{2m_{j}} + \frac{m_{j}\omega_{j}^{2}}{2} \left( \hat{x}_{j} - \frac{c_{j}\hat{\bm V}_v}{m_{j}\omega_{j}^{2}} \right)^{2} \right\}
  \label{eq:bath-Hamiltonian}
\end{align}
and $\hat{\bm V}_v$ represents the system operator involved in the system–bath coupling, defined as
\begin{eqnarray}
  \hat{\bm V}_v = \sum_{n,m} V_{nm}^v(q) \lvert n \rangle \langle m \rvert, 
  \label{eq:VR}
\end{eqnarray}
with $V_{nm}^v(q) = \langle n, q \rvert V^v(\hat q) \lvert m, q \rangle$ for any operator $V^v(\hat q)$.

In HEOM, $\hat{\bm V}_v (q)$ can take an arbitrary form; here, however, we focus on the linear case 
\begin{eqnarray}
\hat{\bm V}_v (q)= q \otimes \bm{I}.
\end{eqnarray}
In this situation, the environmental noise follows a Gaussian distribution, and
the model becomes consistent with the central limit theorem, which states that
random noise converges to a Gaussian distribution regardless of the underlying
mechanism of noise generation.\cite{T06JPSJ} 
It should be noted, however, that in vibrational spectroscopy of solutions,
nonlinear contributions of $\hat{\bm V}_v (q)$ responsible for inhomogeneous broadening must also be
taken into account.\cite{PJT25JCP1,HT26JCP1}

To reduce the computational cost, we employ an Ohmic SDF expressed as
\begin{align}
J^{v}(\omega) = \frac{\hbar \eta_{v}\omega}{\pi},
\label{JOhmic}
\end{align}
where $\eta_{v}$ represents the system-bath coupling strength.
Note that a Brownian system with an Ohmic SDF exhibits an anomaly;\cite{KT24JCP3}
however, since the anomaly is weak, it can be circumvented by appropriate
methods.\cite{IT19JCTC}

\subsubsection{Electromagnetic relaxation}
The atomic electronic states in dissipative environments studied in a former work were modeled using a 3D-RISB framework with rotational symmetry well preserved about the $x$, $y$, and $z$ axes, from which the AO-HEOM were derived.\cite{ZT25JCP1,ZT26JCP1}

For molecular systems, if the dipole components along the x, y, and z directions of nuclear motion remain unchanged, the 3D-RISB model described below can be applied directly. Thus the S-B interaction and bath Hamiltonian in the $x, y$ and $z$ directions is expressed as
\begin{align}
\hat{\bm{H}}_{\mathrm{I+B}}^e = \sum_{\alpha=x,y, z} \hat{H}_{I+B}^{\alpha},
\label{eq:BE}
\end{align}
where
\begin{align}
\hat{H}_{I+B}^{\alpha} = \sum_j \left\{
\frac{(\hat{p}_{j}^{\alpha})^2}{2 m_j^{\alpha}} + \frac{1}{2}m_j^{\alpha}  (\omega_j^{\alpha})^2 \left(\hat{q}_j^{\alpha}  -
\frac{c_{j}^{\alpha} \hat {\bm V}_{\alpha}}{m_{j}^{\alpha} (\omega_{j}^{\alpha})^{2}} \right)^2\right\},
\label{eq:Balpha}
\end{align}
where $ \hat {\bm V}_{\alpha}=  \hat {V}_{\alpha} \otimes \bm{I} $ and 
$m_j^{\alpha}$, $\hat{p}_j^{\alpha}$, $\hat q_j^{\alpha}$ and $\omega_j^{\alpha}$ represent the mass, momentum, coordinate, and frequency, the system part of the S-B interaction, respectively, of the $j$th bath oscillator mode in the $\alpha$ direction.   The system part of the S-B interaction is described as $\hat V_{\alpha}$.  Although the HEOM formalism can accommodate arbitrary forms of $\hat V_{\alpha}$, 
 we adapt $\hat V_{\alpha} = \hat \mu_{\alpha}$, 
corresponding to the standard electromagnetic interaction with a vacuum electric bath, which leads to natural radiation damping. 

For the spin-boson system, the Ohmic SDF is known to exhibit strong infrared
anomalies;\cite{KT24JCP4} therefore, to make the HEOM formalism applicable, we
employ the Drude SDF expressed as\cite{T06JPSJ,T20JCP}
\begin{align}
J^{\alpha}(\omega) = \frac{\hbar \eta_{\alpha}}{\pi}
\frac{\gamma_{\alpha}^{2} \omega}{\gamma_{\alpha}^{2} + \omega^{2}},
\label{JDrude}
\end{align}
where $\eta_{\alpha}$ is the system-bath coupling strength and
$\gamma_{\alpha}$ characterizes the spectral width of the intramolecular bath,
which is inversely related to the vibrational dephasing time
$\tau_{\alpha} = 1/\gamma_{\alpha}$.

If the dipole in the $z$-direction depends on $q$, the operator $V_{z}$ must be
expressed as a function of $q$ and treated in the same manner as the molecular
vibrations discussed in the previous subsection.

\section{MO-HEOM}
\label{sec:MO-HEOM}

MO-HEOM allows for the direct incorporation of thermal bath effects, such as solvents exhibiting three-dimensional rotational symmetry represented by $V_r^{\alpha}$ ($\alpha = x, y, z$), into MO orbitals. For demonstration purposes, we restrict the system to one dimension ($V_v^{z} = V_v$). Although nuclear degrees of freedom can be rigorously treated using the Multistate Low-Temperature Quantum Fokker-Planck equation (MS-LT-QFPE)\cite{IT19JCTC} for general potentials, including dissociative states, many molecular vibrations are harmonic. In such cases, the HEOM with Brownian oscillator SDF (BO-HEOM) framework provides a significant reduction in computational cost.

\subsection{MS-LT-QFPE framework}
\label{sec:MO-LTQFPE}
We introduce the Wigner distribution function, which is the
quantum analogy of the classical distribution function in phase space.
Although computationally expensive, the HEOM in Wigner space is ideal
for studying quantum transport systems, because it allows for the
treatment of continuous systems, utilizing open boundary conditions
and periodic boundary conditions.\cite{TW91PRA,TW92JCP,TM97JCP,MT98CPL} In addition, the
formalism can accommodate the inclusion of an arbitrary time-dependent
external field.\cite{TM97JCP,MT98CPL} Because we can
compare quantum results with classical results obtained in the
classical limit of the HEOM in Wigner space, this approach is
effective for identifying purely quantum effects.
\cite{TW92JCP,ST11JPCA}

For the density matrix $\hat{\bm{\rho}}(t)$, given in
Eq.~\eqref{eq:system-densitymatrix}, the Wigner distribution is defined
by~\cite{TM94JCP,TM97JCP,MT98CPL}
\begin{align}
  \hat{\bm{W}}(p,q;t) &= \sum_{n,m} |n\rangle\, W_{nm}(p,q;t)\,\langle m| ,
  \label{eq:system-Wigner}
\end{align}
where
\begin{align}
  W_{nm}(p,q;t) &\equiv \frac{1}{2\pi\hbar}
  \int \mathrm{d}r\, e^{-ipr/\hbar}\,
  \rho_{nm}\!\left(q+\tfrac{r}{2},\,q-\tfrac{r}{2};t\right).
\end{align}

In terms of the Wigner distribution, the $nm$th element of the quantum
Liouvillian for the system, $\hat{\bm{\mathcal{L}}}_{\mathrm{qm}}\hat{\bm{\rho }}(q,{q}';t)\equiv (i/\hbar )[\hat{\bm{H}}_{\mathrm{S}},\hat{\bm{\rho }}(q,{q}';t)]$, takes the form
\begin{align}
 & \bigl(\hat{\bm{\mathcal{L}}}_{\mathrm{qm}}\hat{\bm{W}}(p,q;t)\bigr)_{nm}
  = \frac{p}{m}\,\frac{\partial}{\partial q} W_{nm}(p,q;t) \notag \\
  &\quad \quad \quad \quad + \frac{i}{\hbar}\sum_{l}\int \mathrm{d}p'\,
      U_{nl}(p-p',q)\, W_{lm}(p',q;t) \notag \\
  &\quad \quad \quad \quad - \frac{i}{\hbar}\sum_{l}\int \mathrm{d}p'\,
      U_{lm}^{\ast}(p-p',q)\, W_{nl}(p',q;t) ,
  \label{eq:quantum-liouvillian}
\end{align}
where 
\begin{align}
  U_{nm}(p,q) &\equiv \frac{1}{2\pi\hbar}
  \int \mathrm{d}r\, e^{-ipr/\hbar}\,
  U_{nm}\!\left(q+\tfrac{r}{2}\right)
\end{align}
are the potentials in Wigner space,
We adopt the [$K_\alpha-1/K_\alpha$] Pad{\'e} approximation, where
$K_\alpha$ is an integer in the $\alpha$ direction, to express the fluctuation and dissipation operators in terms of Pad{\'e} approximated $\nu_k$ and $\eta_k$.\cite{hu2010communication}  

Then the equations of motion are expressed as
\begin{align}
  &\frac{\partial }{\partial t}\bm{W}_{\bf {n}}(p,q,t)  =-\left( \hat{\bm{\mathcal{L}}}_{\mathrm{qm}} +\sum _{k}^{K}n_{k}\nu _{k}+\hat{\Xi }_{K}\right)\bm{W}_{\bf {n}}(p,q,t) \nonumber \\
  &\quad -\sum _{k}^{K}\hat{\Phi }\bm{W}_{{\bf n}+{\bf e}_{k}}(p,q,t) -\sum _{k}^{K}n_{k}\nu _{k}\hat{\Theta }_{k}\bm{W}_{{\bf n}-{\bf e}_{k}}(p,q,t),
  \label{eq:lt-qfpe-d}2
\end{align}
where ${\bf n} \equiv (\dots, n_{k}, \dots)$ is a $K$-dimensional multi-index
whose components are all non-negative integers, and
${\bf e}_{k} \equiv (0, \dots, 1, 0, \dots)$ is the $k$th unit vector.
The multi-index ${\bf n}$ represents the index of the hierarchy, and
physically, the first hierarchical element,
$\bm{W}_{\mathbf{0}}(p,q,t)$, corresponds to the MS-WDF,
$\bm{W}(p,q,t)$.
The remaining hierarchical elements serve only to facilitate the treatment of
the non-Markovian system--bath interaction that arises from quantum
low-temperature effects.

  The operators for the fluctuation and dissipation, $\hat{\Phi }$, $\hat{\Theta }_{k}$, and $\hat{\Xi }_{k}$, appearing in Eq.~\eqref{eq:lt-qfpe-d}, are defined as
  \begin{align}
    \hat{\Phi }&\equiv -\frac{\partial }{\partial p},
  \end{align}
  \begin{align}
    \hat{\Theta }_{k}&\equiv\frac{2\eta _{k}}{\beta \hbar }\frac{\partial }{\partial p},
  \end{align}
and
  \begin{align}
      \hat{\Xi }_{K}&\equiv -\frac{\partial }{\partial p}\biggl(p+\frac{1}{\beta \hbar }\frac{\partial }{\partial p}\biggr) +\sum _{k}^{K}\hat{\Phi }\hat{\Theta }_{k}.
    \label{eq:xi-d}
  \end{align}

Note that for 3D molecules such as methane or benzene, a 3D-RISB model must be employed. In such a model, the reaction coordinate is explicitly decomposed into three spatial components: $r \rightarrow r_z$, $r_x$, and $r_y$. Each bath is coupled to the corresponding 3D component of molecular motion, which can be described either atomistically or in terms of normal modes resolved into 3D elements. The coupling operators $\hat V_v^{\alpha}$ for $\alpha=x, y,$ and $z$ are evaluated in the same way as in the 1D case.

\subsection{BO-HEOM framework}
\label{sec:MO-HEOMBO}
When perturbations from external laser fields or thermal excitation are
sufficiently small, the interatomic distance remains near the thermal
equilibrium configuration of the electronic ground state at $r^{\mathrm{eq}}$.
For a small displacement $\delta q = r - r^{\mathrm{eq}}_0$, the Hamiltonian can
be approximated as
\begin{eqnarray}
  \hat{\bm{H}}_{\mathrm{S}} &&= - \frac{\hbar^2 \nabla_{\delta q}^2 }{2\mu} \otimes \bm{I} + \sum_{n} |n\rangle U_{n}(\delta q) \langle n| \\
   &&+\sum_{m\neq n} |m\rangle \Delta_{mn}\langle n|,
\end{eqnarray}
We express the ground-state potential as  $U_0(\delta q) = {\mu \omega_0^2} \delta q^2/2$ and $\omega_0$ is determined from quantum chemistry calculations. 
The potential surface of the $n$th excited state is approximated as
\begin{eqnarray}
U_n(\delta r) = \frac{\mu \omega_0^2}{2} \left( \delta q - \delta_n \right)^2 + E_n ,
\label{eq:U_n}
\end{eqnarray}
where $\delta_n=r^{\mathrm{eq}}_n-r^{\mathrm{eq}}_0$ denotes the displacement of the oscillator and $E_n$ represents the excitation energy at $r^{\mathrm{eq}}_n$, both of which are determined from quantum chemistry calculations. The ground-state vibrational frequency, $\omega_0$, was assumed to remain unchanged upon electronic excitation.
In this formalism, the coordinate-independent effect of nonadiabatic coupling is represented by $\Delta_{mn}$.

The bath Hamiltonian in this case is given by Eq. \eqref{eq:bath-Hamiltonian} with $\hat{\bm V}_v = \delta q  \otimes \bm{I}$.
Because both the nuclear potential and the bath‑oscillator potential are harmonic, one can introduce suitable normal‑mode coordinates between them and thereby reformulate the system potential plus bath as a single effective bath\cite{Garg1985,TM93PRE,TM94JPSJ,TT09JPSJ,TT10JCP,T12JCP,ST17JPCL}  
\begin{align}
\hat{\bm{H}}_{\mathrm{S}} &+ \hat{\bm{H}}_{\mathrm{I+B}}^v = \hat {\bm H}_S   \nonumber \\
& + \sum_j \left\{
\frac{(\hat{p}_{j}^{v})^2}{2 m_j^{v}} + \frac{1}{2}m_j^{v}  (\omega_j^{v})^2 \left(\hat{q}_j^{v}  -
\frac{c_{j}^{v} \hat {\bm V}'_{v} }{m_{j}^{v} (\omega_{j}^{v})^{2}} \right)^2\right\},
  \label{eq:BOHamiltonian}
\end{align}
where
 \begin{eqnarray}
\hat {\bm H}_S = \sum_n E_n \lvert n \rangle\langle n  \rvert 
\label{eq:HSeigen}
\end{eqnarray}  
and 
\begin{eqnarray}
\hat {\bm V}'_{v} = \sum_n {V}_n' \lvert n \rangle\langle n  \rvert.
\label{eq:VR}
\end{eqnarray}
For Ohmic SDF Eq. \eqref{JOhmic}, we have\cite{TM94JPSJ,TT09JPSJ,TT10JCP,T12JCP,ST17JPCL}
\begin{align}
  J_{\mathrm{BO}} (\omega) = \frac{\hbar\lambda_1}{\pi} \frac{\bar\eta_v \omega_0^2 \omega}{(\omega_0^2 - \omega^2)^2 + \bar\eta_v^2 \omega^2},
  \label{eq:drudePlusBO}
\end{align}
where, the reorganization energy (or Stokes shift) is defined from $\delta_n$ as $
\lambda_n = \frac{\mu \delta_n^2 \omega_0^2}{4}$, and the effective coupling strength between the vibrational mode and the bath is given by $\bar\eta_v = {\eta_v}/{\mu}$.\cite{Garg1985,TM93PRE,TM94JPSJ,TT09JPSJ,TT10JCP}  
For the practical implementation of MO-HEOM, we set  ${V}_n' \equiv \lambda_n/ \lambda_1$.

For this Hamiltonian, we can derive the BO-HEOM expressed as
\begin{eqnarray}
\label{BOHEOM}
\frac{d}{dt} \hat{\rho} _{\{{\bf n}_{\alpha}\}}=&&-\left[ \frac{i}{\hbar}\hat {\bm H}_S^{\times}+ \sum_{k=n_{\rm min}^{v}}^{K_v} \left( n_k^v \nu_k^v \right) \right]\hat{\rho} _{\{{\bf n}_{v}\}} \nonumber \\
&&-\frac{i}{\hbar}\sum_{k=n_{\rm min}^{\alpha}}^{K_v} n_k^v\hat{ {\bm \Theta}}_k^v\hat{\rho}_{\{{\bf n}_{v} - {\bf e}^{k}_{v} \}}  \nonumber \\
&&-\frac{i}{\hbar}\sum_{k=n_{\rm min}^{v}}^{K_v} \hat {\bm V}_{v} ^{' \times} \hat{\rho}
_{\{{\bf n}_{v} + {\bf e}^{k}_{v} \}} , 
\end{eqnarray}
where
${\bf n}_{\alpha}=(n_{n_{\rm min}^{v}}^{v}, n_1^{v}, n_2^{v}, \cdots,   n_{K_{v}}^{v} )$  is a set of integers to describe the hierarchy elements and ${\{{\bf n}_{v} \pm {\bf e}^{k}_{v} \}}$ with the index $k$, where ${\bf e}^{k}_{v} $ is the $k$th unit vector in the $z$ direction.
We defined $\delta_v = \sqrt {\omega_0^2 - {\bar\eta_v^2}/4} $.
The HEOM in Eq.\eqref{BOHEOM} with $n_{\rm min}^{v}=-1$ is then formulated with $\nu_{-1}^v \equiv \bar\eta_v/2+i\delta_{v}$, $\nu_{0}^v \equiv \bar\eta_v/2-i\delta_{v}$.
The operators are defined as 
\begin{equation}
\hat  {\bm \Theta} _{-1}^{v} = \frac{\lambda_v \omega_0^2}{4  \delta_v} 
\left( \hat {\bm V}_{v} ^{' \circ}  - {A}_v^+  \hat {\bm V}_{v} ^{' \times}  \right),
\end{equation}
\begin{equation}
\hat  {\bm \Theta} _{0}^v = \frac{\lambda_v \omega_0 ^2}{4 \delta_v}\left( - \hat {\bm V}_{v} ^{' \circ}  + {A}_v^- \hat {\bm V}_{v} ^{' \times} \right) ,
\end{equation}
and
\begin{equation}
\label{Pade2}
 {\bm \Theta}_{k> 0}^v =-B_v^k \hat {\bm V}_{v} ^{' \times},
\end{equation}
where $\hat{A}^{\circ} \hat{B} \equiv \hat{A} \hat{B} + \hat{B} \hat{A}$, defined for arbitrary operators $\hat{A}$ and $\hat{B}$, and 
\begin{align}
A_v^ \pm  = \frac{2}{\beta\hbar}\left [\frac{1}{i(\bar\eta_v/2 \pm i \delta_v)}+\sum_{k=1}^{K}\frac{2i\eta_k^v (\bar\eta_v/2 \pm  i \delta_v)}{-(\bar\eta_v/2 \pm i \delta_v)^2+(\nu_k^v) ^2}\right ]
\end{align}
and
\begin{align}
{B}_v^k =  \frac{\lambda_r \bar\eta_v \omega_0^2}{{\beta  }}
\frac{2\eta_k^v{\nu} _k^v}{{\left[ {\omega_0^2 + (\nu_k^v) ^2} \right]}^2 - \bar\eta_v^2 (\nu_k^v)^2}.
\end{align}
By introducing the [$K$-1, $K$] type Padé-approximation\cite{YanPade10A} of coth
\begin{equation}
\coth \frac{\beta\hbar\omega}{2}\simeq\frac{2}{\beta\hbar}\left [\frac{1}{\omega}+\sum_{k=1}^{K}\frac{2\eta_k^v \omega}{\omega^2+(\nu_k^v) ^2}\right ],
\end{equation}
where the The parameters $\eta_k^v$ and $\nu_k^v$ denote the Padé-approximated coupling intensity and frequency, respectively. When $K$ increase to $+\infty$, $\eta_k^v$ and $\nu_k^v$ will converge to 1 and ${2k\pi}/{\beta\hbar}$.

In Eq.~\eqref{BOHEOM}, $\hat{\rho}_{\{{\bf n}_{\alpha}\}}$ are hierarchical auxiliary operators, 
with $\hat{\rho}_{\{{\bf n}_{\alpha}=0\}}$ denoting the reduced density operator. 
Since ${\{{\bf n}_{\alpha}\}}$ spans all non‑negative integers, HEOM requires closure expressed as\cite{IT05JPSJ}
\begin{eqnarray}
&&\sum_{k=n_{\rm min}^{v}}^{K_v} \left( n_k^v \nu_k^v \right)\hat{\rho}_{\{{\bf n}_{v}\}} =
-\frac{i}{\hbar}\sum_{k=n_{\rm min}^{v}}^{K_v} n_k^v \hat  {\bm \Theta}_k^v\hat{\rho}_{\{{\bf n}_{v} - {\bf e}^{k}_{v} \}} 
\label{terminator}
\end{eqnarray}
for $\sum_{k=n_{\rm min}^{v}} ^{K_{v}} \left( n_v^k \nu_k \right)\gg \Delta \omega_{\rm max}$, where $\Delta \omega_{\rm max}$ is the largest transition frequency.

\subsection{Including AO-HEOM framework for electromagnetic bath}
\label{sec:AO-HEOM}

Although this increases the computational cost, the effects of the electromagnetic
heat bath described in Eq.~\eqref{eq:Balpha} can be incorporated by embedding the
hierarchical structure of AO-HEOM~\cite{ZT25JCP1,ZT26JCP1} into the hierarchy of
Eq.~\eqref{BOHEOM}. In this formulation, the index set
${\{\mathbf n_v\}}$ is extended to
${\{\mathbf n_\alpha\}} = (\mathbf n_x, \mathbf n_y, \mathbf n_z, \mathbf n_v)$,
together with the corresponding operators $\hat{\Theta}^{\alpha}_k$, etc.

\section{Numerical demonstrations: absorption spectra of H$_2$ molecule}
\label{sec:NumericalDemo}

\subsection{Setting input parameters for MO-HEOM}

We illustrate the procedure by calculating the $H_2$ absorption spectrum, characterized by $N=2$ in Eq.~\eqref{eq:MO}. 
Specifically, we consider the $X^{1}\Sigma_{g}^{+}\!\to\! B^{1}\Sigma_{u}^{+}$ transition, 
which mainly involves the (1$\sigma_g$)(1$\sigma_g$) and (1$\sigma_g$)(1$\sigma_u$) states. 
The potential energy surface (PES) was obtained using the Full Configuration Interaction (FCI) method. 
Results for these two states are summarized in Table~\ref{parameters1},
and the corresponding input parameters in Table~\ref{parameters2}, with further details provided in Appendix~\ref{quantum chemistry part}. 

\renewcommand{\arraystretch}{1.3} 
\begin{table}
    \centering
    \begin{tabular}{||c|c||}
        \hline
        $E_X$ & $-1.172$ \\
        \hline
        $E_B$ & $-0.628$ \\
        \hline
        $\lambda_B$ & 0.00537 \\
        \hline
        $\omega_r$ & 0.02 \\
        \hline
    \end{tabular}
    \caption{\label{parameters2}input parameters for MO-HEOM.}
\end{table}

According to AO-HEOM\cite{ZT25JCP1,ZT26JCP1}, electron relaxation is very weak at the molecular temperature considered, 
so the electronic states are nearly independent. 
We therefore focus only on nuclear relaxation and select $X^{1}\Sigma_{g}^{+}$ and $B^{1}\Sigma_{u}^{+}$ for demonstration. 

The Hamiltonian and interaction operator are defined as
\begin{eqnarray}
H=\begin{pmatrix}
E_X & \Delta_{X-B} \\
\Delta_{X-B} & E_B
\end{pmatrix},
\end{eqnarray}
where $E_A$ and $E_B$ are given in Table~\ref{parameters2}, and $\Delta_{X-B}$ is a term arising when the Born-Oppenheimer approximation breaks down, typically being zero. To facilitate the determination of the initial thermal equilibrium state, we included this term and set $\Delta_{X-B}=0.01$ to ensure that it does not influence either the equilibrium state or the subsequent dynamics. The simulation was then performed under this condition.

\subsection{Absorption spectra}

As a demonstration of MO-HEOM within the BO-HEOM framework, we calculated the absorption spectrum of the $H_2$ molecule.
The procedure for evaluating Eq.~\eqref{eq:R1} is as follows:\cite{T06JPSJ,T20JCP}  
(i) Run the program for Eqs.~\eqref{BOHEOM} and \eqref{terminator} until thermal equilibrium is reached, obtaining $\hat{\rho}_{\{{\bf n}_{v}\}}^{eq}$ as the correlated initial state.  
(ii) Excite the system by the first interaction $\hat{\mu}^{\times}$ at $t=0$.  
(iii) Propagate the perturbed hierarchy elements under HEOM up to time $t$.  
(iv) Compute the response function in Eq.~\eqref{eq:R1} as the expectation value of $\hat{\mu}$.  
(v) Apply a fast Fourier transform to obtain the spectrum.

After setting the parameter values, we executed the MO-HEOM code using Python~3.13.5, 
with CuPy~13.4.1 for CUDA support, NumPy~2.2.5, and SciPy~1.15.3 for special functions and numerical integration. 
The time-dependent HEOM equations were solved with the fourth-order Runge–Kutta method, 
and the transition matrix was computed using SciPy’s \texttt{quad} function, based on the FORTRAN QUADPACK library. 

All computations were performed on a system with an Intel Core i9-13900KF CPU and an NVIDIA GeForce RTX~4090 GPU, 
using CUDA Toolkit~11.8.

\subsection{Results}

\begin{figure}
\centering
\includegraphics[width=1\linewidth]{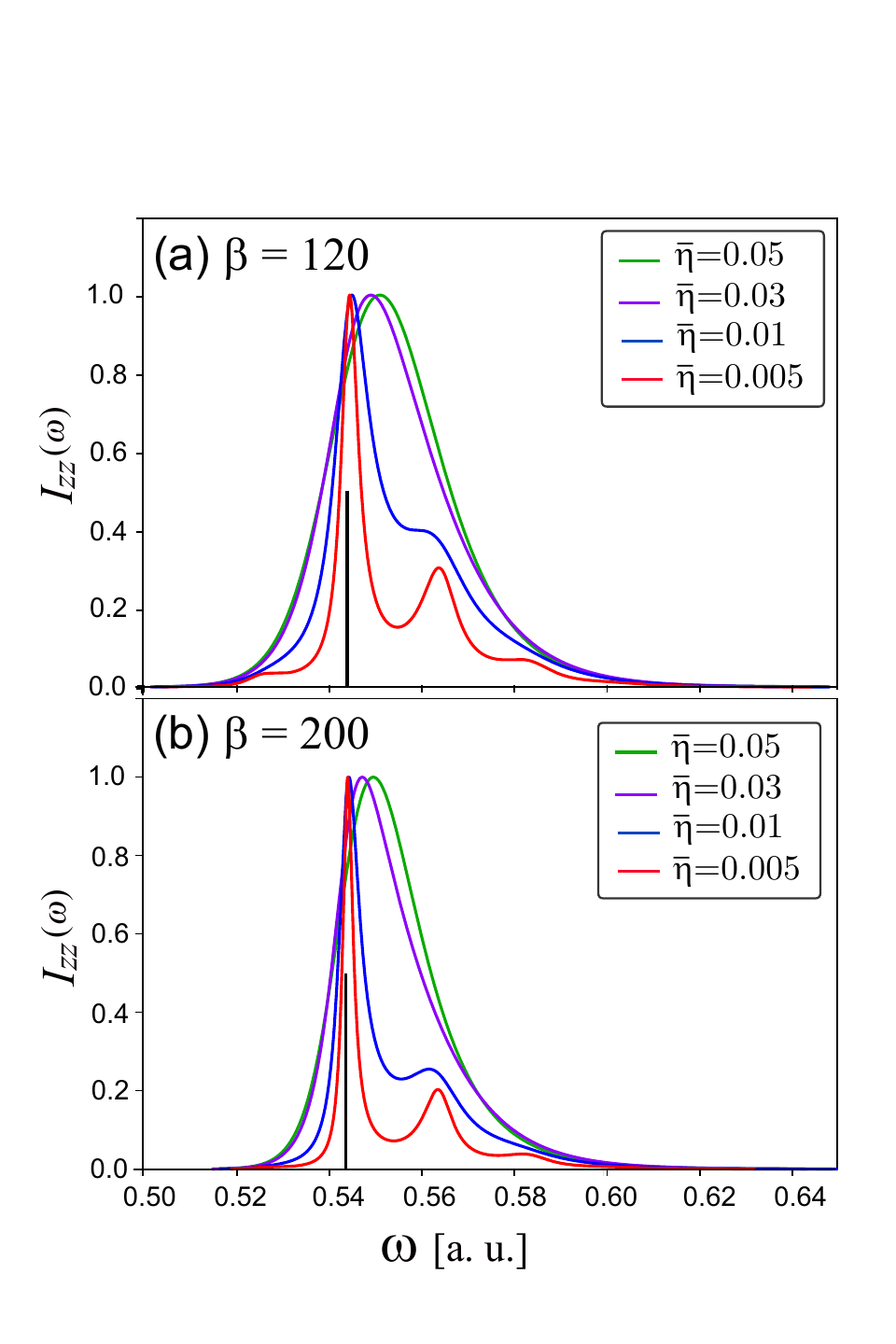}
\caption{\label{absorp_specH2}Linear absorption spectrum of H$_2$ for (a) $\beta = 40$ and (b) $\beta = 60$.
For comparison with the MO-HEOM results, the spectral peaks calculated from the
golden rule [Eq.~\eqref{Goldenrule}] are shown as green vertical lines. Each
spectrum is normalized to its maximum value, whereas the results obtained from
Fermi's golden rule are scaled to 0.5 so as not to obscure the MO-HEOM spectra.
}
\end{figure}

The absorption of the $H_2$ molecule is induced by the dipole operator; for two-level system, we set $\hat \mu=\hat \sigma_x$.  The resulting absorption spectrum is shown in Fig.~\ref{absorp_specH2}. 

For reference, we also depict the absorption spectrum without the bath. From Fermi's golden rule, this is given by
\begin{eqnarray}
I_{\alpha' \alpha }(\omega)=\sum_{mn}{\mu}_{\alpha'}^{mn}{\mu }_{\alpha}^{nm}\frac{e^{-\beta E_m}-e^{-\beta E_n}}{Z} \nonumber \\
\times \delta \left(\hbar \omega+E_m -E_n \right)
\label{Goldenrule}
\end{eqnarray}
where $Z=\sum_n e^{-\beta E_n}$ is the partition function and $\mu_{\alpha}^{mn}=\langle m | \hat \mu_{\alpha} | n \rangle$. 

For inverse temperatures $\beta = 120$ and $200$, corresponding approximately to 2630~K and 1580~K in SI units. We employed the Padé approximations [0/1] and [1/2], and truncated the HEOM at hierarchy level 10 for different values of $\bar\eta_r$, ensuring convergence in all cases.

Following Eq.~\eqref{BOHEOM}, the peak frequencies are associated with energy transitions from the ground state to excited states, accompanied by phonon sidebands characterized by the vibrational frequency $\delta_v = \sqrt{\omega_0^2 - \bar\eta_v^2/4}$.  The vibrational frequency $\delta_v$ is real for $\bar\eta_v < 0.04$, whereas it becomes imaginary for $\bar\eta_v \ge 0.04$.  
Thus, as shown in Fig.~\ref{absorp_specH2}, phonon sidebands are observed for small $\bar\eta_v$, while a featureless single Gaussian peak appears for large $\bar\eta_v$.

Higher temperatures broaden both the electronic transition peak and the vibrational sidebands,
while leaving the vibrational frequency unchanged.
The Golden Rule result exhibits a peak near 0.54, corresponding to the
\(X^{1}\Sigma_{g}^{+} \rightarrow B^{1}\Sigma_{u}^{+}\) transitions without a change in the
vibrational quantum number (\(\Delta n_v = 0\)), since no effective bath is included.
The subsequent vibrational progressions on the left correspond to
\(\Delta n_v = -1 \), whereas those on the right correspond to
\(\Delta n_v = 1, 2,\) and 3.

These results successfully reproduce well-established experimental observations of light absorption. Importantly, they are rigorously derived within the framework of open quantum systems theory, incorporating the direct coupling of spatially structured thermal baths to molecular orbitals.

\section{Conclusion}
\label{sec:Conclution} 

MO-HEOM provides a rigorous, non-perturbative framework for describing quantum thermal fluctuations and dissipation within molecular orbital theory. Once the excited-state density operator and Liouville dynamics are defined, the method yields a fully dynamical description under external fields, naturally extending to nonlinear and multidimensional spectra.\cite{T06JPSJ,T20JCP}

In this work, we restrict our analysis of intramolecular modes to harmonic vibrations. In cases involving strongly anharmonic modes—particularly those associated with nonadiabatic transitions or photodissociation—it is neccesary to treat the relevant degrees of freedom using the MS-LT-QFPE presented in Sec. \ref{sec:MO-LTQFPE}.

While the 3D RISB model can capture spatial anisotropy in relaxation, a more refined description can be achieved by increasing the number of baths as needed: each molecular vibrational mode may be assigned its own bath,\cite{PJT25JCP1,PUT26JCP1} and, if necessary, SDF can be tailored to incorporate inter-bath correlations.\cite{TT20JPSJ}

While introducing thermal baths increases numerical costs, the dissipation effect stabilizes the behavior of wave packets in highly excited states, enabling a reduction in basis functions. The reduced density-matrix formalism is also  expected to integrate with TDDFT, paving the way for hybrid approaches. Thus, MO-HEOM establishes a general foundation for investigating quantum thermodynamic phenomena across spectroscopy, quantum chemistry, and cavity QED, with extensions to multi-electron systems in Fock space.

In the present study, the interaction between the electronic system and the thermal bath is required to preserve rotational symmetry, while still allowing considerable flexibility. Although we assume a thermal-bath coupling that is linear with respect to the molecular axis, more realistic interaction forms can be constructed using machine-learning models trained on trajectories obtained from molecular dynamics (MD) or quantum mechanics/molecular mechanics (QM/MM) simulations.\cite{UT20JCTC,UT21JCTC,PJT25JCP1,PUT26JCP1} However, within the MD or QM/MM frameworks themselves, it is not possible to describe the entanglement between the electronic system and its environment—a defining feature of quantum dissipative systems. The present approach therefore enables quantum-consistent calculations, with the reliability of the results depending solely on the quality of the constructed model, thereby allowing for transparent and well-founded discussion.

\section*{Acknowledgments}
Y. T. was supported by JST (Grant No. CREST 1002405000170).  
Y. Z. is supported by JST SPRING, the establishment of university
fellowships toward the creation of science technology innovation (Grant No. JPMJSP2110). 

\section*{Author declarations}
\subsection*{Conflict of Interest}
The authors have no conflicts to disclose.

\section*{Data availability}
The data that support the findings of this study are available from the corresponding author upon reasonable request.

\appendix
\section{Calculation of the Potential Energy Surface (PES) and Characteristic Frequency of $H_2$}
\label{quantum chemistry part}

In the MO-HEOM with BO framework presented in Sec. \ref{sec:MO-HEOMBO}, model parameters such as $\omega_r$ and $\lambda_r$ are obtained from quantum chemical calculations.
To this end, we first compute the potential energy surfaces (PES) of the ground and excited states.  
The PES represents the total molecular energy under the Born--Oppenheimer approximation, expressed as a function of nuclear coordinates. 
For diatomic molecules, this coordinate is simply the internuclear distance $r$. 
Importantly, the total energy must include both electronic energy and nuclear repulsion, which differs from our previous work.\cite{ZT25JCP1,ZT26JCP1}.

\subsection{Hartree--Fock (HF) Solution}
\label{HF}
In an $N$-electron system, the HF solution is obtained within the single‑determinant approximation and the variational principle. In the linear combination of atomic orbitals (LCAO) representation, a one‑electron state is expressed as \begin{equation} |a\rangle^{\lambda} = \sum_i C_{ia}^{\lambda} \, |\chi_i\rangle^{\lambda}, \end{equation} where $|\chi_i\rangle^{\lambda}$ are atomic basis functions and $C_{ia}^{\lambda}$ are the LCAO coefficients for spin $\lambda=\alpha,\beta$. This orbital is an eigenfunction of the Fock operator, \begin{equation} \hat f^{\lambda} |a\rangle = \epsilon_a^\lambda |a\rangle, \end{equation} with matrix elements 
\begin{align} 
f_{ij}^\alpha &= h_{ij} + J_{ij}^\alpha + J_{ij}^\beta - K_{ij}^\alpha, \\ f_{ij}^\beta &= h_{ij} + J_{ij}^\alpha + J_{ij}^\beta - K_{ij}^\beta. \end{align}
Here,
\begin{align}
h_{ij}&=\langle i|-\tfrac{1}{2}\nabla^2-\sum_A \tfrac{Z_A}{|r-R_{A}|}|j\rangle, \label{eq:core hamiltonian}\\
J^\lambda_{ij}&=\sum_{x\in \text{occ}}\sum_{kl}C_{k x}^{\lambda}C_{l x}^{\lambda}\langle ik|jl\rangle,\\
K^\lambda_{ij}&=\sum_{x\in \text{occ}}\sum_{kl}C_{k x}^{\lambda}C_{l x}^{\lambda}\langle ik|lj\rangle,
\end{align}
where the electron repulsion integral is defined as
\begin{equation}
\label{ao-eri}
\langle ij|kl\rangle=\iint dr_1 dr_2 \, 
\chi_i(r_1)\chi_j(r_2)\frac{1}{|r_1-r_2|}\chi_k(r_1)\chi_l(r_2).
\end{equation}

For convenience, indices $ijkl$ denote AO basis functions, while $abcd$ denote HF basis functions.  
We used \texttt{psi4}\cite{10.1063/5.0006002} for the calculations, specifying molecular geometry, charge, multiplicity, basis set, and HF type.  
Nuclear repulsion energy is extracted via \texttt{wfn.molecule().nuclear\_repulsion\_energy()}.

\subsection{Post-HF Treatment}
\label{PHF}
The HF solution is based on the single-determinant assumption, which does not correctly account for electron correlation.  
Thus, post-HF methods are required. Configuration Interaction (CI), especially Full CI (FCI), provides accurate ground and excited state energies and is considered the gold standard.

For $H_2$, a two-electron system, the Hamiltonian can be constructed analogously to helium.  
Matrix elements involving different spin states, specifically unbarred orbitals $d$ corresponding to $\alpha$ spin and barred orbitals $\bar d$ corresponding to $\beta$ spin, are expressed as
\begin{equation}
\label{two electron states}
\langle a \bar{b} | \hat{O}|c\bar{d}\rangle
= \delta_{ac}\langle \bar{b}|\hat{O}_1|\bar{d}\rangle
+ \delta_{\bar{b}\bar{d}}\langle a|\hat{O}_1|c\rangle
+ \langle a \bar{b}|\hat{O}_2|c \bar{d} \rangle,
\end{equation}
where $\hat{O}_1$ and $\hat{O}_2$ denote one- and two-electron operators, respectively, written in HF orbitals.

The one-electron and two-electron matrix elements are
\begin{equation}
\label{one electron operator}
\langle a|\hat{O}_1|c\rangle = \sum_{ij} C_{ia} C_{jc} \langle i|\hat{O}_1|j\rangle,
\end{equation}
\begin{equation}
\label{two electron operator}
\langle a \bar{b}|\hat{O}_2|c \bar{d} \rangle
= \sum_{ijkl} C_{ia} C_{j\bar{b}} C_{kc} C_{l\bar{d}} \langle i j|\hat{O}_2|k l \rangle.
\end{equation}

WWe employed \texttt{psi4.core.MintsHelper.ao\_eri}, \texttt{wfn.Ca}, and \texttt{wfn.H} to extract the two-electron integrals [Eq.~\eqref{ao-eri}], the HF coefficients $C_{ia}^\lambda$, and the core Hamiltonian [Eq.~\eqref{eq:core hamiltonian}], respectively.  
The Hamiltonian matrix was then diagonalized using \texttt{numpy.eigh} to obtain the FCI energies for both singlet and triplet states, following the procedure described in our previous work.\cite{ZT26JCP1}  
The total molecular energy was computed as the sum of the post-HF electronic energy and the nuclear repulsion energy.

\subsection{Global and Local PES Calculation}
\label{PES}
The above procedure is performed for fixed nuclear positions.
Since the HF solution depends on the nuclear coordinates, varying the internuclear distance
requires recalculation from the beginning.
Because constructing the CI Hamiltonian is computationally expensive, density fitting for
evaluating electron–repulsion integrals and Davidson methods for obtaining a limited number of
eigenvalues are often employed to compute approximate low-lying PES at reduced cost.

Thus, we first compute the PES over a wide range using density fitting and Davidson methods to
locate local minima, and then apply FCI only to several points near these minima to obtain
accurate results. Demonstration code for calculating global and local \(H_2\) PES using
\texttt{numpy} and the \texttt{psi4} library will be provided.

\begin{figure}
\centering
\includegraphics[width=1\linewidth]{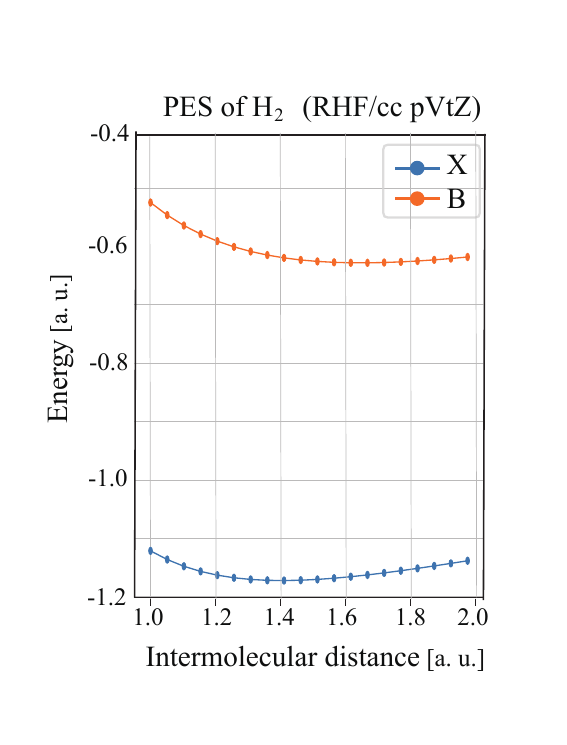}
\caption{\label{H2PES} The potential energy surfaces of the $\Sigma_{g}^{+}$ and $B^{1}\Sigma_{u}^{+}$ states, given in atomic units (a.u.).}
\end{figure}

The local PES of the \(X^{1}\Sigma_{g}^{+}\) and \(B^{1}\Sigma_{u}^{+}\) states, as part of the
calculated results, are plotted in Fig.~\ref{H2PES}.

\subsection{Characteristic Frequency from Local PES}
\label{charafreq}
Near a local minimum, the PES can be approximated by a quadratic function, and the nuclei behave as harmonic oscillators.  
For a two-body system, motion can be separated into total and relative motion, with masses
\begin{equation}
M=m_1+m_2,\quad \mu=\frac{m_1m_2}{m_1+m_2}.
\end{equation}
Since the PES of $H_2$ depends on the relative coordinate, the potential is written as
\begin{equation}
U(r)=\tfrac{1}{2}\mu \omega^2 (r-r^{eq})^2,\quad
\omega=\sqrt{\tfrac{2A}{\mu}},
\end{equation}
where, $\mu$ is the reduced mass, and $A$ is the quadratic coefficient optimized from the ground- and excited-state potential energy surfaces (PESs).

In many standard quantum chemistry programs, vibrational frequencies are routinely computed from the Hessian matrix evaluated at equilibrium geometries. However, as of this writing, only ground-state frequencies are implemented in \texttt{Psi4}. As an alternative, we fit the PES data to a quadratic function using \texttt{numpy.polyfit}.

The PES of the \(X^{1}\Sigma_{g}^{+}\) state is fitted as  
\(0.2012(r - 1.4100)^2 - 1.1724\), using data points 6–11 (counted from the left),  
whereas the \(B^{1}\Sigma_{u}^{+}\) state is fitted as  
\(0.0902(r - 1.6411)^2 - 0.6281\), using points 11–20.  
All quantities are given in atomic units (a.u.).

The reduced mass in atomic units is \(\mu = 918.075\), and the corresponding vibrational
frequencies are calculated to be 0.0209 and 0.0140. The displacement between the equilibrium
positions is \(\delta_{X-B} = r_B^{\mathrm{eq}} - r_X^{\mathrm{eq}} = 0.2311\), and the
reorganization energy is 
\[
\lambda_{X-B} = \mu\,\delta_{X-B}^2\,\omega_X^2/4 = 0.00537.
\]

More accurate results can be obtained using much larger basis sets.~\cite{SILKOWSKI2021255}

\renewcommand{\arraystretch}{1.3} 
\begin{table}
    \centering
    \begin{tabular}{||c|c|c||}
        \hline
        & $X^{1}\Sigma_{g}^{+}$ & $B^{1}\Sigma_{u}^{+}$ \\
        \hline
        $r^{eq}$ & 1.4100 & 1.6411\\
        \hline
        $E$ & -1.1724 & -0.6281\\
        \hline
        $\omega$ & 0.0209 & 0.0140\\
        \hline
    \end{tabular}
    \caption{\label{parameters1}Parameters from quantum chemistry results (all quantities are given in atomic units, a.u.).}
\end{table}

\subsection{The variational method and expansions for larger molecules}

The present pilot FCI calculations for the H$_2$ molecule complete instantaneously. However, for general polyatomic systems with realistic basis sets, the exponentially scaling nature of FCI renders it impractical.  
For larger systems, a practical alternative is to truncate the CI expansion. The coupled-cluster (CC) method reformulates the FCI wavefunction using an exponential ansatz, and, in general, truncated CC methods yield more accurate results than truncated CI methods at the same level of excitation.

However, standard CC theory provides only the ground-state wavefunction. Methods for computing
excited-state energies were developed later, such as equation-of-motion coupled-cluster
(EOM-CC),\cite{Emrich1981a,Emrich1981b,Sekino1984,Geertsen1989} coupled-cluster linear
response (CC-LR),\cite{Ghosh1981,Takahashi1986,Koch1990a,Koch1990b,Rico1993} and
symmetry-adapted-cluster configuration interaction (SAC-CI),\cite{Nakatsuji1981,Hirao1981}
all of which extend CC theory to excited states based on the ground-state CC solution.
The truncation errors of EOM-CC at various levels have been compared with FCI in previous
studies.\cite{Hirata03} Beyond truncated CI/CC methods, the frozen-core approximation may also be employed when
appropriate.

For polyatomic molecules, the normal-mode frequencies are obtained by diagonalizing the
mass-weighted Hessian matrix at a local minimum, which is typically located using a
quasi-Newton optimization method.

\bibliography{tanimura_publist,AO-HEOM,references,MO-HEOM}

\end{document}